# High power 1640-nm Er:$Y_2O_3$ ceramic laser at room temperature

Hangbin Xie,[1] Jianing Zhang,[1,3] Fei Wang,[2] Deyuan Shen,[1,2,*] Jun Wang,[1,2] and Dingyuan Tang[1,2]

[1]*Jiangsu Key Laboratory of Advanced Laser Materials and Devices, School of Physics and Electronic Engineering, Jiangsu Normal University, Xuzhou 221116, China*

[2]*Jiangsu Collaborative Innovation Center of Advanced Laser Technology and Emerging Industry, Jiangsu Normal University, Xuzhou 221116, China*

[3]*e-mail: zhangjianing@jsnu.edu.cn*

*Corresponding author: mrde@jsnu.edu.cn*

**Abstract:** We report on high power operation of Er:$Y_2O_3$ ceramic laser at ~1.6 μm using low scattering loss, 0.25 at.% $Er^{3+}$ doped ceramic sample fabricated in-house via co-precipitation process. The laser is in-band pumped by an Er, Yb fiber laser at 1535.6 nm and generates 10.2 W of continuous-wave (CW) output power at 1640.4 nm with a slope efficiency of 25% with respect to the absorbed pump power. To the best of our knowledge, this is the first demonstration of ~1.6 μm Er:$Y_2O_3$ laser at room temperature. The prospects for further scaling in output power and lasing efficiency via low $Er^{3+}$ doping and reduced energy-transfer upconversion are discussed.

Solid-state laser sources operating in the eye-safe wavelength range of ~1.6 μm have a variety of applications including remote sensing, laser radar, free-space communications, and as the pump source for generation of mid-infrared (3–5 μm) radiation via nonlinear frequency conversion [1,2]. Typically, laser sources at this wavelength region can be obtained directly by pumping solid state laser gain media doped with $Er^{3+}$ ions. Although Yttrium Aluminum Garnet (YAG) remains a most widely used host material for solid state lasers owing to its unique thermal-mechanical and optical properties [3,4], alternative hosts with superior thermal-mechanical and spectral properties suitable for laser generation of specific requirements have long been the research focus of the laser community.

Cubic sesquioxides ($Y_2O_3$, $Sc_2O_3$ and $Lu_2O_3$) have drawn recently considerable attention as laser gain hosts due to their superior thermal conductivity and relatively low phonon energy that facilitate laser generation of longer wavelengths, opening up the prospects for high power ~3-μm laser operation [5,6]. However, fabrication of large size and high quality sesquioxide single crystal is still challenging today due to their relatively high melting points [7,8]. With the developments of fabrication technology, high quality sesquioxide ceramics with various RE doping have been successfully fabricated at temperatures much lower than their melting points and efficient laser operation at different wavelengths have been demonstrated [9-11].

In-band pumped at ~1.5 μm ($^4I_{15/2}\rightarrow{^4I_{13/2}}$), room temperature laser oscillation of $Er^{3+}$ doped $Lu_2O_3$ and $Sc_2O_3$ single crystal at ~1.6 μm has been successfully demonstrated with output power 0.7 W and 0.95 W respectively [12,13]. Benefit from improved thermo-optic properties at reduced temperature, Er:$Sc_2O_3$ ceramic laser generated 3.3 W of 1558 nm output with 45% of slope efficiency with liquid nitrogen cooling [14]. Cryogenically-cooled $Er^{3+}:Y_2O_3$ ceramic laser of 0.5 at.% doping has generated 9.3W of 1.6 μm output with 65% slope efficiency when in-band pumped at 1536 nm, and up to 24 W of 1.6 μm and 13 W of 2.7 μm simultaneously with overall efficiency of 62% when pumped at 980 nm and with a cascade laser design [15,16].

In this paper, we report on room temperature high power operation of home-developed low scattering loss $Er:Y_2O_3$ ceramic prepared via co-precipitation process. A relatively low $Er^{3+}$ doping of 0.25 at.% was used to alleviate the energy transfer upconversion (ETU) processes and guarantee room temperature operation. Pumped using a 1535.6 nm Er, Yb fiber laser directly to the upper energy level $^4I_{13/2}$, the laser generated up to 10.2 W of CW output power at 1640.4 nm for 41.9 W of absorbed pump power at 1535.6 nm, corresponding to a slope efficiency of 25% with respect to the absorbed pump power. To the best of our knowledge, this is the first demonstration of room temperature $Er^{3+}$ doped sesquioxide ceramic laser at ~1.6 μm.

The pump source is a home-constructed high power Er, Yb fiber laser with 30 μm diameter (0.2 NA) of Er, Yb co-doped phosphor-silicate core surrounded by a 350 μm diameter (0.49 NA) pure silica inner-cladding. A volume Bragg grating (VBG) of ~0.4 nm bandwidth and > 99% reflectivity at the center wavelength of 1570 nm was employed for wavelength control. The emission wavelength of the pump fiber laser was tuned to match the absorption peak of the $Er:Y_2O_3$ ceramic at 1535.6 nm by adjusting the incident angle of VBG. The fiber pump source, having ~2.2 of $M^2$ factor and 0.2 nm of bandwidth (FWHM), was collimated and focused to a beam diameter of ~300 μm in $Er:Y_2O_3$ ceramic. The confocal parameter of the pump light inside $Er:Y_2O_3$ was estimated to be ~79 mm.

A simple two-mirror resonator was employed for laser characterization as shown in Fig. 1. The resonator comprised a plane pump input coupler (IC) with high reflectivity (R > 99%) coated at the lasing wavelength (1600-1700 nm) and high transmission (T > 98%) coated at 1500-1550 nm, a 100 mm radius-of-curvature concave output coupler (OC) of 2% or 5% transmission ($T_{oc}$) at the lasing wavelength and high reflectivity (R > 99%) at the pump light. A 45° dichroic mirror with high reflectivity (R > 99%) at the lasing wavelength and high transmission (R < 1%) at 1535 nm was used to separate the ~1.6 μm laser and the residual pump light after the OC. $Er:Y_2O_3$ ceramic sample of 0.25 and 0.5 at.% $Er^{3+}$-doping were prepared using co-precipitation process. With optimized preparing processes, intragranular pores in the ceramic were greatly diminished and hence reduced scattering losses and improved optical homogeneity. The ceramic sample was cut and polished to have a dimension of 2×3 mm$^2$ in cross section and 18 mm in length, and both end faces were AR-coated at 1500-1700 nm. Single-pass small signal absorption of the 0.25 and 0.5 at.% doped ceramic at 1535.6 nm was measured to be ~75% and ~94% under non-lasing condition. The sample was wrapped with indium foil and mounted on a water-cooled copper heat-sink maintained at a temperature of 15 °C to allow for efficient heat removal. The physical length of the laser resonator was ~36 mm and the IC was placed as close as possible to the Er:Y$_2$O$_3$ ceramic. Output power and spectrum were recorded using a power meter (Ophir) and an optical spectrum analyzer of 50 pm resolution (AQ6375B, Yokogawa). The beam quality of the Er:Y$_2$O$_3$ ceramic laser was characterized using a beam profiler (NanoScan, Photon Inc.). Upconversion emission spectrum was recorded using a spectrometer of 1 nm resolution (HR4000, Ocean Optics).

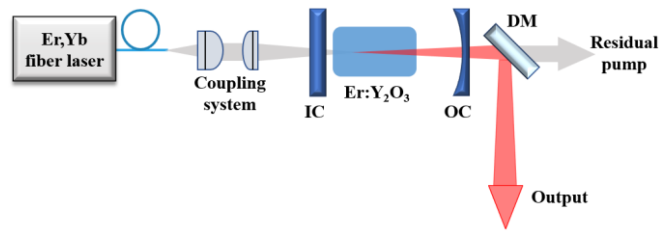

**Fig. 1.** Experimental setup of the in-band pumped Er:Y$_2$O$_3$ ceramic laser.

Room temperature lasing behavior of the Er:Y$_2$O$_3$ ceramic were first characterized using output couplers of different transmissions. The 0.25 at.% doped sample reaches threshold at a pump power of ~0.86 W and ~1.64 W, and generates 5.2 W and 3.9 W of ~1.64 μm output power at 21.8W of absorbed pump power for the T$_{oc}$=2% and 5% coupler, respectively. The corresponding slope efficiency are 24.9% and 19.8% (as shown in Fig. 2). For the 0.5 at.% sample, we did not get lasing using either the 2% or 5% OC, this should be attributed to the relatively severe ETU effect of Er:Y$_2$O$_3$ [17].

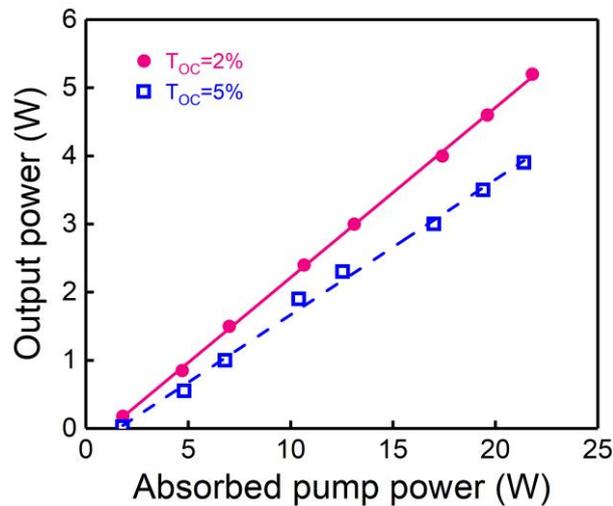

**Fig. 2.** Output power versus absorbed pump power for output couplers of 2% and 5% transmission.

Energy level diagram and room-temperature upconversion emission spectrum of the 0.25 at.% doped Er:Y$_2$O$_3$ ceramic under 1.53 μm excitation is shown in Fig. 3. It can be seen that $^2H_{11/2}$, $^4S_{3/2}\rightarrow{}^4I_{15/2}$ green emission (centered at ~550 nm) dominates the spectrum, the $^4F_{9/2}\rightarrow{}^4I_{15/2}$ red (660~683 nm) and $^4I_{11/2}\rightarrow{}^4I_{15/2}$ ~980 nm emission are relatively weak. No blue emission observed from $^2H_{9/2}$ to the ground state due to the relatively low Er$^{3+}$ concentration and hence reduced ETU effect. The strong green emission can be explained by the relative long lifetime of the $^4I_{9/2}$ energy level (~0.1 ms) that enhances population of the high lying $^2H_{11/2}$, $^4S_{3/2}$ level via excited state absorption (ESA) process [18]. The moderate lasing efficiency demonstrated at room temperature can as well be attributed to the relatively strong upconversion processes that deplete the upper laser level via ETU and ESA processes. Further reduction in $Er^{3+}$ doping concentration should be a practical way to reduce the ETU effect and improve lasing performance of room temperature Er:$Y_2O_3$ laser at ~1.6 μm.

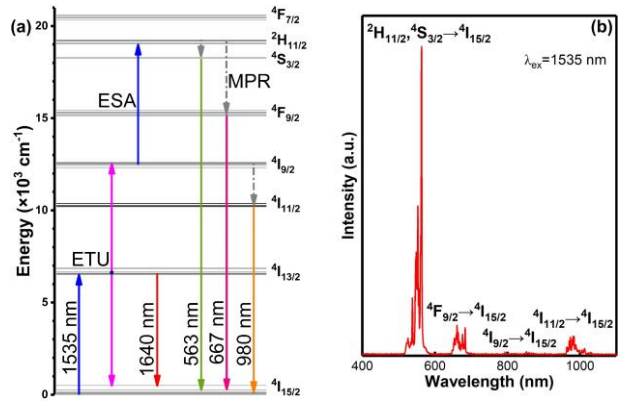

**Fig. 3.** Energy level diagram (a), and fluorescence spectrum (b) of the 0.25 at.% Er:$Y_2O_3$ ceramic excited at 1535.6 nm.

Figure 4 shows high power operation of the 0.25 at.% Er:$Y_2O_3$ using a output coupler of 2% transmission. The laser generated 10.2 W of output power for ~41.9 W of absorbed pump power, corresponding to a slope efficiency of 25%. It should be noted that output power increases linearly with pump power indicating that there is scope for further scaling in output power with improved pump. The inset of Fig. 4 shows a typical output spectrum of the Er:$Y_2O_3$ ceramic laser. The spectrum centered at 1640.4 nm and with a wavelength span of ~1 nm.

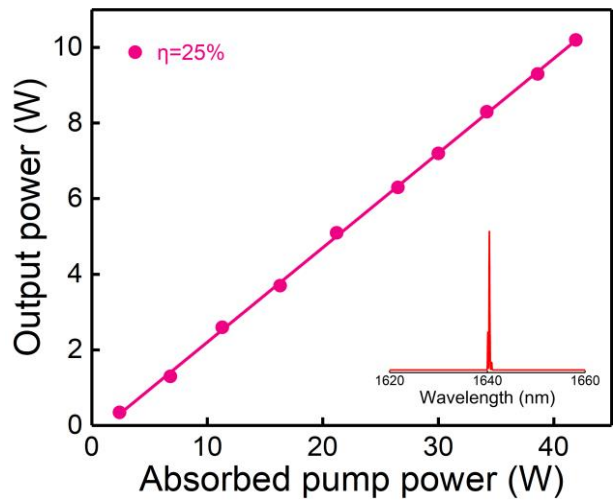

**Fig. 4.** Output power versus absorbed pump power with a OC of T=2%. The inset shows a typical output spectrum of the Er:$Y_2O_3$ ceramic laser.

Placing a plano-convex lens of 100 mm focal length behind the output coupler to focus the laser beam to a beam profiler, the M² factor was measured to be ~1.05 and ~1.08 in the horizontal and vertical directions respectively, and the inset shows the profile of the laser beam near the focus with typical Gaussian distribution, as shown in Fig. 5.

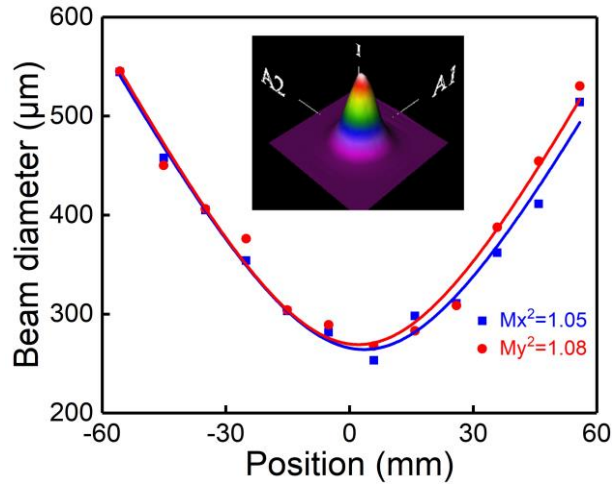

**Fig. 5.** Beam quality factor measurement of the Er:$Y_2O_3$ laser. The inset shows the transverse 3D beam profile.

In conclusion, we demonstrated for the first time to the best of our knowledge room temperature operation of an Er:$Y_2O_3$ ceramic laser at ~1.6 μm using home-developed low scattering loss Er:$Y_2O_3$ ceramic of 0.25 at.% doping. The laser generated 10.2 W of 1640 nm output power for ~41.9 W of absorbed pump power at 1535 nm, corresponding to a slope efficiency with respect to the absorbed pump power of 25%. Further improvement in lasing efficiency and output power should be possible employing Er:$Y_2O_3$ ceramic of lower $Er^{3+}$ concentration and hence reduced upconversion effects.

**Funding**. This work is supported by the National Natural Science Foundation of China (NSFC) under contract No. 62035007, 61875078 and in part by the Postgraduates Innovation Program of Jiangsu Normal University (2020XKT770).